# Brain Ageing Prediction using Isolation Forest Technique and Residual Neural Network (ResNet)

Saadat Behzadi[1], Danial Sharifrazi[2], Roohallah Alizadehsani[2,3], Mojtaba Lotfaliany[4], and Mohammadreza Mohebbi[3]

[1]Department of Electronic Engineering, University of Bologna, Bologna, Italy
[2]Institute for Intelligent Systems Research and Innovations (IISRI), Deakin University, Geelong, Australia
[3]Biostatistics Unit, Faculty of Health, Deakin University, Geelong, Australia
[4]The Institute for Mental and Physical Health and Clinical Translation (IMPACT), School of Medicine, Deakin University, Geelong, Australia

**Abstract.**

Brain aging is a complex and dynamic process, leading to functional and structural changes in the brain. These changes could lead to the increased risk of neurodegenerative diseases and cognitive decline. Accurate brain-age estimation utilizing neuroimaging data has become necessary for detecting initial signs of neurodegeneration. Here, we propose a novel deep learning approach using the Residual Neural Network 101 Version 2 (ResNet101V2) model to predict brain age from MRI scans. To train, validate and test our proposed model, we used a large dataset of 2102 images which were selected randomly from the International Consortium for Brain Mapping (ICBM). Next, we applied data preprocessing techniques, including normalizing the images and using outlier detection via Isolation Forest method. Then, we evaluated various pre-trained approaches (namely: MobileNetV2, ResNet50V2, ResNet101V2, Xception). The results demonstrated that the ResNet101V2 model has higher performance compared with the other models, attaining MAEs of 0.9136 and 0.8242 years for before and after using Isolation Forest process. Our method achieved a high accuracy in brain age estimation in ICBM dataset and it provides a reliable brain age prediction.

**Keywords:** Brain Aging, Isolation Forest, Residual Neural Network (ResNet), Deep LearningFirst Section

## 1 Introduction

Brain Aging is an ongoing process defined by neuroanatomical and biological changes occurring over time, resulting in a progressive Deterioration in the physical and mental functionality of the brain. Such changes consist of structural alterations in the brain, including changes in specific brain regions, reduced brain volume, and loss of neural connections [1,2]. While the changes related to brain aging are not necessarily pathological, the risk of cognitive deterioration and neurodegenerative diseases increases with age [3,4]. However, the impact of aging on the brain could be significantly different among individuals. As a result, according to the current knowledge, boosting our understanding of brain aging is crucial for better detecting early neurodegeneration and predicting age-related cognitive decline.

In recent decades, numerous studies have focused on estimating brain age using magnetic resonance imaging (MRI). T1-weighted MRI, among various MRI modalities, has shown a superior performance for brain age estimation as this type effectively depicts the brain's structure and the morphological changes related to aging [5-7]. As a result, it has become the preferred option in most studies for estimating brain age [8].

Cole et al investigated the use of convolutional neural networks (CNN) to estimate brain age as a biomarker of brain aging. They used a large dataset of healthy individuals, demonstrating that from both raw T1-weighted MRI and pre-processed data, CNN could be able to precisely predict chronological brain age, with MAE of 4.16 and 4.66 years for pre-processed data and raw data respectively [9]. Moreover, Peng et al. developed a Simple Fully Convolutional Network (SFCN) for brain age estimation, as they believed this model is more suitable for small datasets. To achieve high performance, techniques such as data augmentation, pre-training, and prediction bias correction were employed in this article. This model demonstrated strong performance, achieving a mean absolute error (MAE) of 2.14 years [10]. Hong et al. also introduced a ten-layer 3D CNN to predict the brain age of healthy children aged 0 to 5 years, achieving high performance with a MAE of 67.6 days. Due to the small number of training datasets, data augmentation techniques such as image rotation, gamma correction, and scaling were applied to expand the training data [11]. Liu et al. introduced a multi-feature-based network (MFN) for brain age estimation. In their work, six types of morphological features—cortical thickness, surface area, cortical volume, folding index, curvature index, and local gyrification index—were used to construct the MFN. The model achieved a MAE of 3.73 years [12].

Unlike global brain area methods, Popescu et al. introduced a U-Net model as a deep learning framework to estimate brain age across localized brain regions, achieving the most precise predictions in the prefrontal cortex and periventricular regions, with a MAE of approximately 7 years [13]. Jiang et al. used CNNs with images of seven brain networks to develop age estimation models. The CNN models attained the most precise results for the default mode network (DMN), dorsal attention network (DAN), and frontoparietal network (FPN), with MAEs of 6.07, 5.55, and 5.77 years, in that order. For the other four networks, the MAEs were more than 8 years [14].

In addition to deep learning models, traditional machine learning techniques have also been widely utilized for brain age prediction, such as Gaussian process regression (GPR), relevance vector regression (RVR), and support vector regression (SVR) [15-19]. For instance, Cole et al. developed a GPR model by creating a similarity matrix from 3D T1-weighted MRI scans, which were segmented into gray and white matter, normalized, and concatenated into vectors. This matrix was used as input for the GPR model to predict brain age. The results demonstrated that the GPR model could accurately estimate brain age, achieving an MAE of 7.08 years on the test dataset from the Lothian Birth Cohort [15]. Franke et al. used the RVR method to estimate brain age, achieving a MAE of 5 years in healthy adults [16]. They also used the RVR method to predict brain age in children and adolescents. They utilized MRI images acquired from 394 healthy children and adolescents. This method showed a high performance, with an MAE of 1.1 years [17]. Moreover, using data from healthy individuals, Gaser et al. applied the Brain-AGE approach based on RVR to enhance the detection of progression from mild cognitive impairment to Alzheimer's disease [18]. Liem et al. extracted feature vectors from connectivity matrices, cortical thickness, cortical surface area, and subcortical volumes to develop a brain age estimation model using SVR as the initial stage and Random Forest (RF) as the subsequent stage. Their findings showed that all models achieved high performance, with MAEs ranging from 4.29 to 7.29 years [19]. In addition, Ganaie et al used an improved twin support vector regression (ITSVR) model to predict brain age. ITSVR, when tested on MRI data from cognitively healthy individuals, patients with mild cognitive impairment, and those with Alzheimer's disease, achieved a MAE of 2.77 years [20].

Despite the fact that recent research has concentrated on estimating brain aging, this is an emerging field and there are still several challenges in improving the accuracy and reliability of these models. One major challenge is enhancing regression accuracy in predicting brain age. Another significant hurdle is the performance of deep learning algorithms, especially when applied to a small dataset. While deep learning has shown its effectiveness in many fields, some studies have shown that using deep learning algorithms does not necessarily outperform simpler machine learning approaches, such as SVM or Linear regression. In other words, it remains unclear whether using more complex deep learning models leads to substantial improvements. As a result, further research is required to deepen our understanding.

In this paper, a proposed methodology is presented to predict brain age. The structure of the method is based on using several pre-trained models, including ResNet101V2, ResNet50V2, MobileNetV2 and Xception, to find the best model to use in the modeling training section. In addition, to improve accuracy and minimize the amount of error, an outlier detection method is used. To achieve this goal, the Isolation Forest method is applied. The proposed method achieved a state-of-the-art result in brain age prediction, recording MAEs of 0.9136 and 0.8242 years for before and after using the Isolation Forest method. These results shows that the proposed methodology outperforms various commonly used machine learning models reported in the literature. The code related to the proposed method in available at the following GitHub link: https://github.com/danialsharifrazi/Brain-Age-Prediction-by-Outlier-Detection-and-Transfer-Learning-.

The rest of the work is organized as follows: first, a brief discussion of the dataset is given. Then, the main steps of the proposed method for brain age estimation are described. In the next section, the performance and results of the proposed approach are reported. Finally, a brief discussion and comparison with other state-of-the-art methods are done and some conclusions are included.

## 2    Subjects (Dataset)

All neuroimaging data utilized in this study, consisting of T1-weighted MRI scans, were gathered from International Consortium for Brain Mapping (ICBM, https://www.loni.usc.edu/ICBM/) database [21]. The dataset is collected from healthy individuals, with a gender distribution of 673 males and 673 females (male/female rate: 1/1 and total: 1346). The ages of the participants range from 18 to 80 years, with a mean age of 36.21 and standard deviation of 14.82. Figure 1 displays examples of the neuroimaging data used in this study.

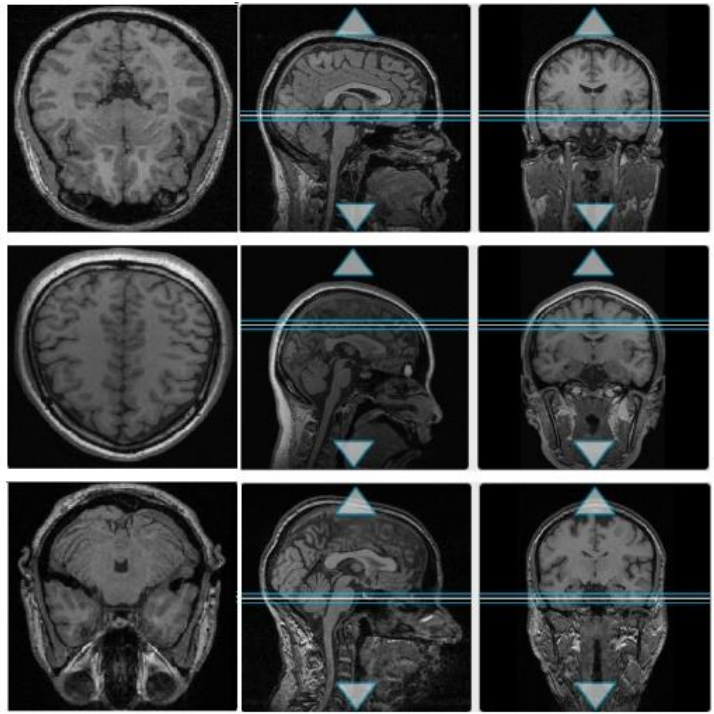

**Figure 1.** Some examples of the input images utilized in this study.

## 3    Proposed Methodology

In the proposed method, the ICBM dataset is used to estimate the brain age of people based on their brain images. In the paper method, due to hardware limitations, we use 2102 different images randomly to provide us with a suitable modelling of the entire dataset. In this dataset, images are stored in MNC, DCM formats. Firstly, all images were converted to PNG format to be a processable format for the OpenCV package.

After converting the images format to PNG, the size of the images will be 224×224×3. To use images with Transfer Learning methods, all images must be converted to this size to input the model. Nibabel and Pydicom packages were used to convert images from MNC and DCM format to PNG, respectively [22].

In the next stage, all images are normalized between zero and one. In the proposed method, 5-fold Cross-Validation method is used to ensure the correctness of evaluation metrics. It is used to divide the data into training and test data. The proposed method is executed 5 times, and the average errors are reported as the final error. Also, all data are used to divide the data into training and test data.

Next, outlier data should be identified. Low quality images can have a significant impact on training the final model. For this purpose, we use the Isolation Forest method. The number of dataset images before outlier detection was 2102 images. This number reduced to 1988 images after using the outlier detection method. In this way, 114 non-trainable images were detected for the proposed model. So, they were removed from the dataset.

The ResNet101V2 method is used in the model training section. This model is a pre-trained model that we can give our own images to. In the training section, other pre-trained models were also used to identify the best model. In this regard, ResNet50V2, MobileNetV2 and Xception models were used as well.

To train the ResNet101V2 model, ImageNet weights were set to the model. Then we use a Global Average Pooling layer to reduce the size of the output images. The next layer is a fully connected layer (Dense layer) with 1024 neurons. Having an extra layer at the end of the model can level up the model training with new data.

Since the output of the model is a number that is related to the age of individual's brain, there is a one-neuron layer as the output layer in the proposed model. In this method, we set the initial layers of the model as "Trainable=False" and set the last 10 layers of the model as "Trainable=True" so that in addition to the fully connected layer at the end, the last 10 layers of the model can also be trained with new images. Other model parameters are shown in Table 1. Keras package with Tensorflow backend is used to implement the pre-trained models.

Finally, the test data is given to the model and the model makes its prediction from the brain age of the people. Then the evaluation metrics are calculated according to the actual age of the people and the age predicted by the models. The results of this research show that the proposed method is able to predict the brain age with a very low error. The implementation process of the proposed method is described step by step below. Also, Figure 2 shows the proposed method in a brief and step-by-step manner.

0. Start
1. Convert images (format and size)
2. Pre-process images (Normalization between 0 and 1)
3. Outlier detection (Isolation Forest)
4. Split the dataset (5-fold Cross-Validation)
5. Trian the model with our dataset (ResNet101V2)
6. Predict the ages (test phase)
7. Calculate evaluation metrics
8. End

Table 1. Hyper-parameters of the proposed method.

| Hyper Parameters | Value |
|---|---|
| Loss | Mean Squared Error (MSE) |
| Optimizer | Adam |
| Learning Rate | 0.001 |
| Epochs | 20 |
| Weights | Image Net |
| Bath Size | 128 |
| Activation Function | Relu |
| Input Size | 224*224*3 |
| Output Size | 1 |

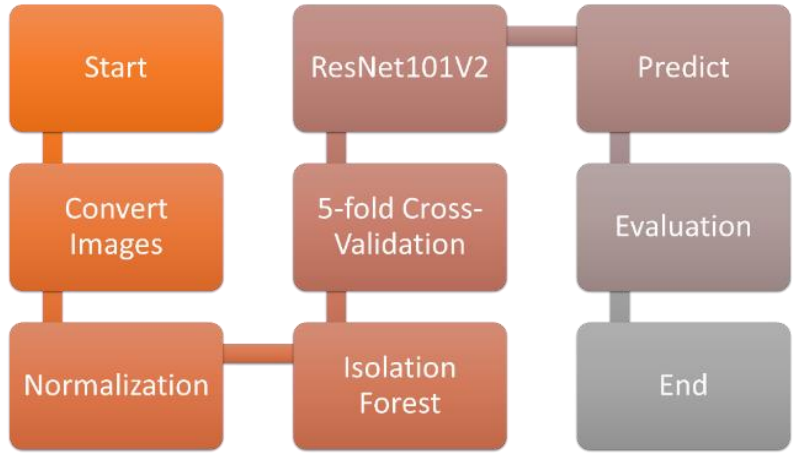

Figure 2. Proposed method in brief.

## 4    Experimental Results

As previously noted in the article, the proposed method demonstrated strong performance. Table 2 illustrates the evaluation metrics for the proposed method based on the mean of 5 folds before using Isolation Forest method for four different pretrained methods. Based on the results, the accuracies of ResNet50V2, ResNet101V2, and Xception methods were better than that of MobileNetV2 with MAEs of 1.1438, 0.8242, and 0.9106 years respectively, compared to MobileNetV2's MAE of 87.5886. In the standard process (before using the outlier detection), the best performance was related to ResNet101V2 with the lowest error.

Table 3 shows the evaluation metrics for the proposed method based on the mean of 5 folds after using Isolation Forest method. Compared to the results in the table 3, higher performance was achieved here. Similar to before using Isolation Forest method, the results related to ResNet101V2 were better than other methods, with an MAE of 0.8242, compared with the MAEs of 87.5886, 1.1438, and 0.9106 for MobileNetV2, ResNet50V2, and Xception respectively. Moreover, to compare the performance of standard method with that of after using Isolation Forest method, figure 3, plotted on a logarithm scale, shows the MAEs related to both approaches.

**Table 2.** Evaluation matrices for the proposed method based on the mean of 5 folds before using the outlier detection.

| Method | MSE | RMSE | MAE | MAPE | Training Time |
|---|---|---|---|---|---|
| MobileNetV2 | 12166.7012 | 110.0918 | 109.7952 | 3.402758633 | 20.8212 |
| ResNet50V2 | 2.3875 | 1.5314 | 1.3655 | 0.0425 | 36.122 |
| ResNet101V2 | 1.2876 | 1.1294 | 0.9136 | 0.0274 | 54.8855 |
| Xception | 2.3182 | 1.4808 | 1.1849 | 0.035 | 47.6882 |

**Table 3.** Evaluation matrices for the proposed method based on the mean of 5 folds after using the outlier detection.

| Method | MSE | RMSE | MAE | MAPE | Training Time |
|---|---|---|---|---|---|
| MobileNetV2 | 7798.5071 | 87.8257 | 87.5886 | 2.7515 | 19.7347 |
| ResNet50V2 | 1.8331 | 1.3255 | 1.1438 | 0.0362 | 34.8845 |
| ResNet101V2 | 1.0634 | 1.0214 | 0.8242 | 0.0254 | 53.0558 |
| Xception | 1.43 | 1.1935 | 0.9106 | 0.0276 | 43.9202 |

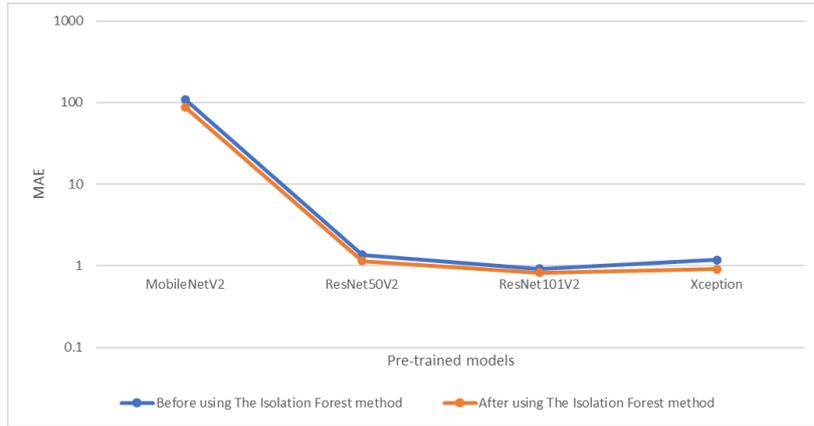

**Figure 3.** Comparison of MAEs before and after applying the Isolation Forest method.

## 5      Discussion

This section highlights some other state-of-the-art methods to provide a clearer comparison with the proposed method. It is worth mentioning that the other leading methods used different datasets, and none focused exclusively on ICBM dataset. So, other research papers are listed in the table to provide a better insight into recent trends in brain age estimation using machine learning algorithms.

For a direct comparison, Peng et al. used a SFCNN to predict brain age. Their results show that they obtained one of the highest accuracies of brain age prediction. Peng et al. used UK Biobank data with 14503 subjects and they obtained an MAE of 2.17 years. Moreover, Ganaie et al utilized a ITSVR model to predict brain age. To train their model, the used 976 subjects and their model obtained a high performance, with an MAE of 2.77 years. In comparison, our proposed methodology seems to attain better performance related to these two methods, with an MAE of 0.8242 years after using the Isolation Forest method. However, it should be noted that the number of dataset images that we used in our model was 2102 from ICBM, which was different from these two mentioned methods.

Quite similar to the size of our dataset images, Liu et al worked with a sample of 2,501 healthy individuals collected from several public databases. They utilized MFN for brain age estimation. Their model achieved a MAE of 3.73 years. Cole et al. also conducted a study using a sample of 2001 individuals acquired from the Brain-Age Normative Control (BANC) dataset. They applied a CNN model, achieving an MAE of 4.16 years. In addition, Popescu et al. used brain MRI scans from 3,463 healthy individuals to train a U-Net model. In this model, they recorded a median mean absolute error of 9.5 years. Here also, the results of our proposed method demonstrate higher performance in predicting brain age. For better comparison, Table 4 outlines a brief summary of the brain age prediction accuracies reported in several published research in recent years. Also, a list of abbreviations used in Table 4 is provided in the footer.

One of the main reasons behind the proposed method's excellent performance is employing effective preprocessing technique within its algorithm. Clearly, data preprocessing is essential in machine learning methods. In addition to this reason, it can be mentioned that using pretrained methods and outlier detection are effective to achieve better performance.

**Table 4.** Recorded results on brain age estimation in literature.

| Articles | Modalties | Input data | Methods | Subjects (Age range) | r | MAE | R2 |
|---|---|---|---|---|---|---|---|
| Cole et al. [9] | [1]sMRI | GM volume map | CNN | 2001 NC (18–90 years) | 0.96 | 4.16 years | 0.92 |
| Peng, et al. [10] | sMRI | 3D minimally-preprocessed T1 brain images | SFCN | 14,503 NC (44-80 years) | - | 2.14 years | - |
| Hong, et al. [11] | T1-weighted MRI | 2D T1-weighted images | 3D CNN | 220 children (0–5 years) | 0.98 | 67.6 days | - |
| Liu, et al. [12] | sMRI | T1-weighted MRI (6 morphological features) | (MFN) | 2501 (20-94 years) | - | 3.73 years | - |
| Popescu, et al. [13] | sMRI | Brain MRI scans (3D maps) | U-Net | 3,463 NC (18–90 years) | - | 9.5 years (median) | - |
| Jiang et al. [14] | sMRI | Functional connectivity; Structural measures | CNN, GPR, RVR | 1454 NC (18–90 years) | 0.87 | 5.55 years | 0.76 |
| Cole et al. [15] | sMRI | GM+WM volumes | GPR | 2001 NC (18–90 years) | 0.94 | 5.02 years | 0.88 |
| Franke et al. [16] | sMRI | GM volume | RVM | 655 NC (19–86 years) | 0.92 | 5.00 years | - |
| Liem et al. [19] | rs-fMRI + sMRI | Functional connectivity; Structural measures | SVR+RF | 2354 NC (19–82 years) | - | 4.29 years | - |
| Ganaie, et al. [20] | sMRI | T1-weighted MRI | (ITSVR) | 976 NC (18–96 years) | - | 2.77 years | 0.97 |

sMRI: Structural Magnetic Resonance Imaging, GM: Gray Matter, CNN: Convolutional Neural Network, NC: Normal Controls (healthy subjects), SFCN: Spatial Fully Convolutional Network, MFN: Morphological Feature Network, GPR: Gaussian Process Regression, RVR: Relevance Vector Regression, FPN: Frontoparietal Network, DAN: Dorsal Attention Network, DMN: Default Mode Network, SMN: Somatomotor Network, VAN: Ventral Attention Network, VN: Visual Network, LN: Limbic Network, WM: White Matter, RVM: Relevance Vector Machine, rs-fMRI: Resting-State Functional Magnetic Resonance Imaging, SVR: Support Vector Regression, RF: Random Forest, ITSVR: Improved Twin Support Vector Regression

## 6     Conclusion

In this study, we have developed a proposed methodology based on neuroimaging data that can estimate brain age with high accuracy. We trained, validated and tested our proposed method by using 2102 different images which were chosen randomly from the ICBM dataset. Besides, to guarantee the accuracy of the evaluation metrics, we utilized the 5-fold Cross-Validation method. Our results revealed that our proposed approach can provide efficient and reliable performance for estimating brain age. It can also be used as a biomarker to assist in diagnosing neurocognitive disorders. In reality, the link between brain age and chronological age is not straightforward. In the future, we will examine some other deep learning and nonlinear models to achieve and provide a more precise prediction of brain age.